\documentclass{emulateapj}

\begin{document}

\title{SCP\,06F6: A carbon-rich extragalactic transient at redshift
  $z\simeq0.14$?}

\author{Boris T. G\"ansicke, Andrew J. Levan, Thomas R. Marsh, Peter J. Wheatley}
\affil{Department of Physics, University of Warwick, Coventry CV4 7AL,
UK}
\email{Boris.Gaensicke@warwick.ac.uk}


\begin{abstract}
We show that the spectrum of the unusual transient SCP\,06F6 is
consistent with emission from a cool, optically thick and carbon-rich
atmosphere if the transient is located at a redshift of
$z\approx0.14$. The implied extragalactic nature of the transient
rules out novae, shell flashes, and V838\,Mon-like events as cause of
the observed brightening. The distance to SCP\,06F6 implies a peak
magnitude of $M_I\simeq-18$, in the regime of supernovae.  While the
morphology of the light curve of SCP\,06F6 around the peak in
brightness resembles the slowly evolving Type~IIn supernovae SN\,1994Y
and SN\,2006\,gy its spectroscopic appearence differs from all
previous observed supernovae. We further report the detection of an
X-ray source co-incident with SCP\,06F6 in a target of opportunity
\textit{XMM-Newton} observation made during the declining phase of the
transient. The X-ray luminosity of
$L_\mathrm{X}\simeq(5\pm1)\times10^{42}\,\mathrm{ergs\,s^{-1}}$ is two
orders of magnitude higher than observed to date from supernovae. If
related to a supernova event, SCP\,06F6 may define a new class.  An
alternative, though less likely, scenario is the tidal disruption of a
carbon-rich star.
\end{abstract}

\section{Introduction}
Studies of local and distant supernovae continue to reveal broad
diversity in supernova properties, both spectral and photometric.
This includes the recognition of SN with extremely high ejecta
velocities (hyper-novae, e.g. \citealt{galamaetal98-1,
  mazzalietal02-1, hjorthetal03-1}) and those which reach peak
magnitudes markedly brighter than seen previously
\citep{smithetal07-2, quimbyetal07-1}. Indeed, even the ``standard
candle" SN\,Ia include examples of unusually discrepant SN, e.g.  the
extremely faint ($M_B=-15.9$) SN\,2007ax
\citep{kasliwaletal08-1}. This variety most likely reflects similar
variations in progenitor properties such as mass, metallicity and
rotation.  Given the dramatically increased rate of SN detection over
the past twenty years (there were 20 reported in 1987, 163 in 1997 and
572 in
2007\footnote{http://cfa-www.harvard.edu/iau/lists/Supernovae.html}),
it is perhaps unsurprsing that the classification system has required
adaptation. Nonetheless the discovery of supernovae whose properties
are broadly different from those seen before remains rare, and it is
these examples which could perhaps place the strongest constraints on
unusual processes in stellar evolution. In fact, from the theoretical
perspective, it appears that the full variety of supernovae has not
been discovered, e.g. \citet{bildstenetal07-1} suggest the existence
of faint ($M_V=-15$ to $-18$) thermonuclear supernovae from helium
cataclysmic variables, or the tidal disruption and ignition of a white
dwarf by intermediate mass black holes \citep{rosswogetal08-1,
  rosswogetal09-1}.

It is clear that the growing number of large-area imaging surveys with
high temporal cadence (hours to days), and the next generation of wide
field space based instruments, will dramatically increase our
knowledge about rare transient events. Projects that are currently
operating, or coming online in the foreseeable, span a wide range of
aperture sizes, e.g. ASAS \citep{pojmanski97-1} or SuperWASP
\citep{pollaccoetal06-1} with limiting magnitudes $\sim13-15$, the
Catalina Real-Time Transient Survey (CRTS, \citealt{drakeetal08-1}) or
SkyMapper \citep{kelleretal07-1} with limiting magnitudes of
$\sim19-21$, PanSTARRS \citep{hodappetal04-1} and LSST
\citep{ivezicetal08-1} with limiting magnitudes of $\sim24-25$, and
ultimately SNAP, which may reach $\sim 28$th magnitude
\citep{aldering05-1}.

Examples of very unusual events serendipitously discovered by deep
supernova surveys are a red transient in M85 [possibly the the result
of a stellar merger \citep{kulkarnietal07-1} but see
\citealt{thompsonetal08-1} for an alternative], and the recent optical
transient SCP\,06F6 reported by \citet{barbaryetal08-1}, which is so
far of unknown nature.

Here, we suggest that SCP\,06F6 had an extragalactic nature with a
redshift of $z\simeq 0.14$, and may represent a sofar unknown type of
supernova or, less likely, a tidal disruption event.

\section{The unusual transient SCP\,06F6}

\citet{barbaryetal08-1} discovered the optical transient SCP\,06F6 as
part or the \textit{Hubble Space Telescope} Cluster Supernova
Survey. The object reached a peak magnitude of $i_\mathrm{775}\simeq
z_\mathrm{850}\simeq21$, and showed roughly symmetric rise and decay
times of $\simeq60$\,d each. No counterpart is detected down to
$i_\mathrm{775}\simeq26.4$ and $z_\mathrm{850}\simeq26.1$. The optical
spectra of SCP\,06F6 obtained with Keck and the VLT were relatively
red, peaking at $\sim6100$\,\AA, and contained several broad
($\sim200-300$\,\AA) absorption troughs blue-wards of
$\sim6500$\,\AA. \citet{barbaryetal08-1} discussed the possible nature
of SCP\,06F6 on the basis of its unusual spectral appearance, but were
not able to find any fully convincing solution. In particular, they
cross-correlated the spectrum of SCP\,06F6 against the spectral
database of the Sloan Digital Sky Survey (SDSS), and noted that the
best match was found with broad absorption line quasars (BAL QSOs) and
carbon-atmosphere (DQ) white dwarfs. The spectra of DQ white dwarfs
with temperatures in the range $\simeq6000-10000$\,K contain broad
absorption bands from C$_2$ \citep{dufouretal05-1, koester+knist06-1},
also known as Swan bands, which are roughly equally spaced in
wavelength. However, \citet{barbaryetal08-1} noted that the positions
of these bands in DQ white dwarfs from SDSS did not line up with the
absorption features in the spectrum of SCP\,06F6.  DQ white dwarfs
display a rich variety in the general morphology of the Swan bands,
however, the position of these absorption troughs remains largely
constant \citep{harrisetal03-1}. Furthermore, the spectral energy
distribution of DQ white dwarfs is bluer than that of SCP\,06F6.

\begin{figure}
\includegraphics[width=\columnwidth]{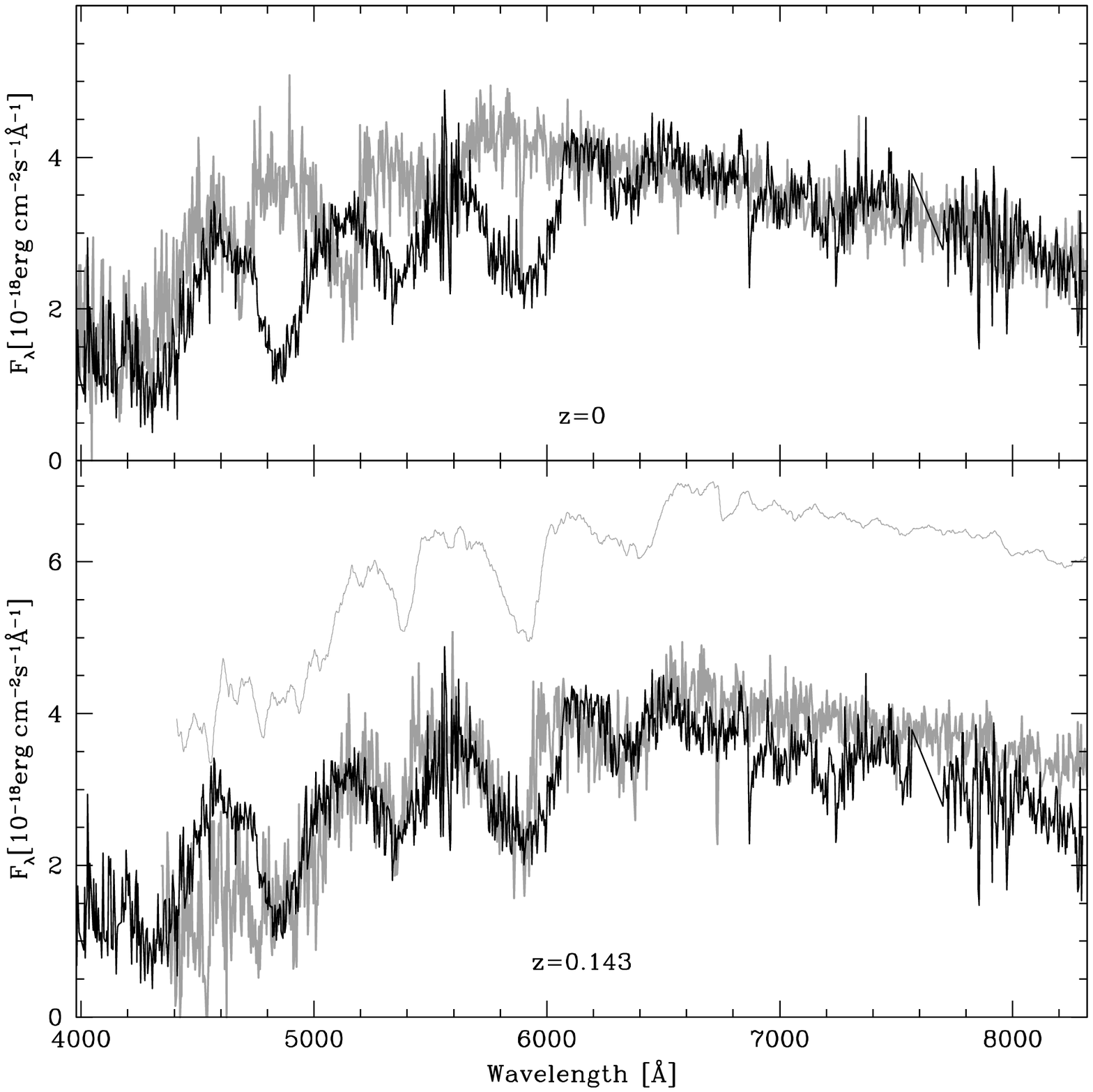}
\caption{\label{f-spectrum} Top panel: the spectrum of SCP\,06F6 (thin
  black line, co-added from the Keck and VLT data of
  \citealt{barbaryetal08-1}) and the SDSS spectrum of the carbon star
  SDSS\,J001836.23-110138.5 \citep{downesetal04-1}. While the spectral
  energy distribution of the two spectra are similar, the positions of
  the C$_2$ Swan bands totally disagree. Bottom panel: same as before,
  but the spectrum of the carbon star has been redshifted to
  $z=0.143$, which brings the positions of the Swan bands in agreement
  with the absorption troughs seen in SCP\,06F6. Shown as a thin gray
  line offset by 2.6 flux units is the carbon star spectrum convolved
  with a  $4000\,\mathrm{km\,s^{-1}}$ outflow velocity
  profile. This illustrates that an expanding envelope will smooth out
  to some extent the sharp C$_2$ band-heads.}
\end{figure}

\section{A carbon-rich optically thick atmosphere in SCP\,06F6}
Upon inspection of the spectrum of SCP\,06F6, we noticed a striking
resemblance to carbon stars, in particular to several of the faint
high-latitude carbon stars presented by \citet{margonetal02-1}. In
order to quantify this resemblance, we carried out a spectral template
fitting to the spectrum of SCP\,06F6.

We obtained the Keck and VLT spectra of SCP\,06F6
\citep{barbaryetal08-1} from the Supernova Cosmology Project web
page. Both spectra were co-added on the wavelength grid of the VLT
spectrum to improve the signal-to-noise ratio. We then created a
template library from the SDSS spectra of 251 carbon stars identified
by \citet{margonetal02-1} and \citet{downesetal04-1}.  The templates
were binned to the wavelength grid of the VLT spectrum, and normalised
to the spectrum of SCP\,06F6 over the range $4000-5500$\,\AA. The
quality of the fits was evaluated by calculating $\chi^2$ over the
wavelength range 4000--7500\,\AA, as well as by visual inspection of
each fit.  

A first run through the template library confirmed the finding of
\citet{barbaryetal08-1} that the positions and spacings of the C$_2$
Swan bands do not coincide with the absorption troughs seen in the
spectrum SCP\,06F6 (Fig.\,\ref{f-spectrum}, top panel).

We then redshifted the carbon star template spectra to adjust the
position of their strongest Swan band with the broad absorption
feature centered at $\simeq5800$\,\AA\ in the observations of
SCP\,06F6. A redshift of $z=0.143\pm0.005$ resulted in  good
agreement in the wavelengths of all the broad absorption features.
The best match in fitting the template library with $z=0.143$ to
the spectrum of SPC\,06F6 is found for the carbon star
SDSS\,J001836.23-110138.5, which reproduces  the spectral features
and the overall spectral shape of SCP\,06F6 well (Fig.\,\ref{f-spectrum},
bottom panel).

The presence of C$_2$ in the spectrum of SCP\,06F6 implies a
relatively low temperature ($\sim5000-6000$\,K) of the emitting
region, consistent with the red spectral energy distribution.
One galactic transient is known that exhibited a spectrum similar to
that of SCP\,06F6: Sakurai's Object (V4334\,Sgr,
\citealt{pavlenkoetal00-1}). Sakurai's object was explained as the
final helium flash of a hydrogen-deficient post-asymptotic giant
branch star, resulting in an optically thick carbon-rich expanding
pseudo-photosphere \citep{duerbeck+benetti96-1, herwig01-1}.  However,
the redshift of $z\simeq0.14$ implies that SCP\,06F6 reached an
absolute magnitude $M_I \sim -18$, which rules out a physical nature
similar to Sakurai's Object, for which $M_V>-4$
\citep{miller-bertolami+althaus07-1}. Similarly, red luminous
variables, such as V838\,Mon or M31-RedVar ($M_\mathrm{bol}\sim-10$;
\citealt{richetal89-1}), or classical novae ($M_V\simeq-9$;
\citealt{cappaciolietal89-1}) can be excluded.

We conclude that the emission of SCP\,06F6 observed during peak
brightness originated in a cool, optically thick and carbon-rich
atmosphere. We stress, however, that this morphological result does
not imply that the progenitor was a normal carbon star~--~just as the
progenitor of Sakurai's Object was not a carbon star.

\begin{figure}
\includegraphics[angle=-90,width=\columnwidth]{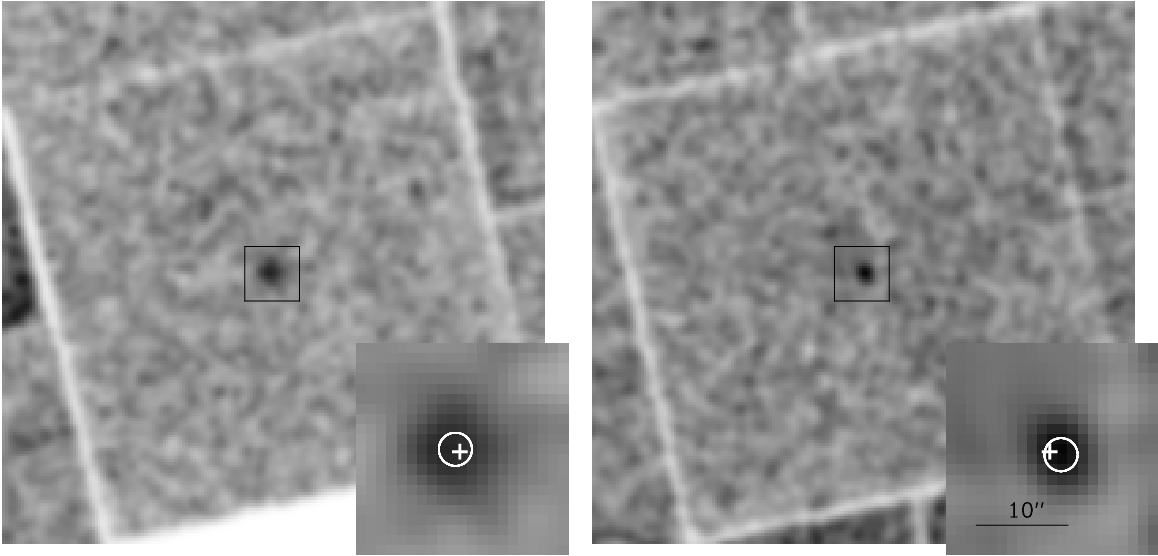}
\caption{\label{f-xmm} X-ray images of SCP\,06F6 in the energy range
  0.2--2.0\,keV taken simultaneously with the two EPIC MOS instruments
  on XMM-Newton (left: MOS1, right: MOS2). Inset panels show the
  expected source position (marked with a cross) and the $1\,\sigma$
  error on the detected X-ray source position, adopting the mean
  uncertainty for faint sources in the 2XMM catalogue 
  \citep{watsonetal09-1}.}
\end{figure}

\section{X-ray detection of SCP\,06F6}
In addition to the optical observations reported in
\citet{barbaryetal08-1}, we have also examined a target of opportunity
\textit{XMM-Newton} observation of SCP\,06F6 made on 2nd August 2006
during the declining phase of the transient's light curve.  The 15\,ks
observation was severely affected by a high radiation background,
which forced premature ends to the exposures with the EPIC
cameras. The resulting images have high background levels and exposure
times of 10, 8 and 3\,ks respectively in the MOS1, MOS2 and PN cameras.
An X-ray source is clearly detected in the soft X-ray images of both
EPIC MOS cameras, and within the positional error the location of the
X-ray source is consistent with that of SCP\,06F6 (Fig.\,\ref{f-xmm}).
The MOS1 count rate for this source in the 0.2--2.0\,keV band is
0.0138$\pm$0.0038\,s$^{-1}$, corresponding to a flux of
$1\times10^{-13}\,\mathrm{ergs\,s^{-1}cm^{-2}}$ in the same band.  At
a redshift of $z=0.143$ this would correspond to an X-ray luminosity
of $L_\mathrm{X}=(5\pm1)\times10^{42}\,\mathrm{ergs\,s^{-1}}$.

The position of SCP\,06F6 has also been observed at two epochs with
the \textit{Chandra} X-ray observatory, once before the transient (1st
April 2003) and once after the \textit{XMM-Newton} pointing (4th
November 2006). No source is detected at the position of SCP\,06F6 at
either epoch. A flux upper limit of
$7.8\times10^{-15}\,\mathrm{ergs\,s^{-1}cm^{-2}}$ in the 0.5-7.0\,keV
band is reported by \citet{barbaryetal08-1} for the first epoch.  We
have examined the 5\,ks observation from the second epoch and derive a
99\% confidence upper limit to the count rate of 0.00084\,s$^{-1}$,
corresponding to a robust flux upper limit of
$1\times10^{-14}\,\mathrm{ergs\,s^{-1}cm^{-2}}$ in the 0.5--7.0\,keV
band.  For all reasonable thermal and non-thermal model spectra, the
flux in the 0.2--2.0\,keV band must be below
$6\times10^{-15}\,\mathrm{ergs\,s^{-1}cm^{-2}}$.

We conclude that the \textit{XMM-Newton} X-ray detection of SCP\,06F6
is associated with the transient outburst, and that the X-ray
luminosity of this object must have increased and then declined again
by at least an order of magnitude with respect to any quiescent
emission.

\begin{figure}
\includegraphics[width=\columnwidth]{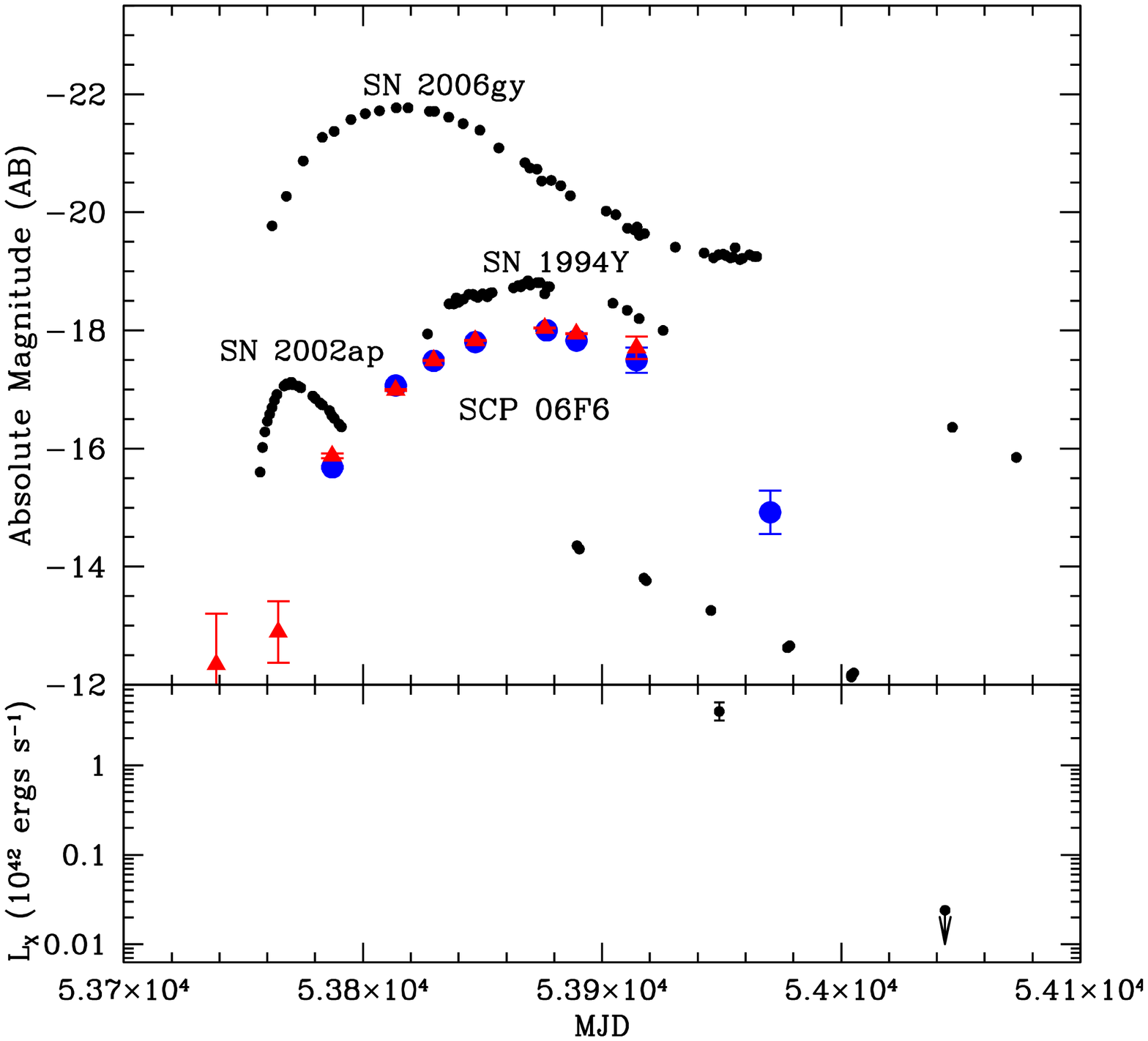}
\caption{\label{f-lc} Top panel: the light curve of SCP\,06F6,
  assuming its redshift to be $z=0.14$ (blue circles $i_{775}$, red
  triangles $z_{850}$), it reaches a peak at both I and Z of $M_{I,Z}
  \sim -18$. Also shown for comparison are the light curves of SN
  2002ap (Ic; \citealt{gal-yametal02-1}), SN\,2006gy (IIn;
  \citealt{smithetal07-2}) and the slowly evolving SN\,1994Y (IIn;
  \citealt{hoetal01-1}). The points have been plotted relative to the
  time frame of SCP\,06F6 and thus the location of other SN points on
  the time axis is arbitrary. Bottom panel: X-ray luminosities of
  SCP\,06F6 implied by the \textit{XMM-Newton} and \textit{Chandra}
  observations. In addition, the object was not detected in a
  \textit{Chandra} observation obtained $\sim1000$~days before the
  onset of the optical outburst, consistent with an upper limit of
  $L_\mathrm{X}\simeq2\times10^{39}\,\mathrm{erg\,s^{-1}}$. }
\end{figure}

\section{Discussion}
In the two previous sections, we concluded that the morphology of the
optical spectrum of SCP\,06F6 at maximum brightness indicates an
origin in a cool, optically thick and carbon-rich atmosphere at a
redshift of $\simeq0.14$, and that the optical transient was
accompanied by a luminous X-ray transient. Here, we discuss the
implications for the nature of SCP\,06F6.

The symmetric light-curve of SCP\,06F6 prompted
\citet{barbaryetal08-1} to consider micro-lensing as an explanation,
although, as they pointed out, the large minimum amplification
($>120$) and the $\sim 100$ day timescale are hard to understand on
such a hypothesis. It can be added that the light-curve does not have
the extended wings of microlensing light curves. If our identification
of Swan bands is correct, then the microlensing hypothesis seems even
more implausible. If microlensing was the cause, then the source would
be a carbon star at $z\simeq0.14$ (dismissing the unlikely combination of
a new type of cool transient \textit{and} microlensing).  The
brightest carbon stars have absolute magnitudes around $-4$
\citep{rebeirotetal93-1}, and so microlensing would have to provide at
least 14 magnitudes or a factor of $400,000$ amplification. Roughly
the right timescale and amplification result from a
$5\,\mathrm{M}_\odot$ object, placed half-way to the carbon star,
passing within $4\,\mathrm{AU}$ of the line-of-sight to the carbon
star at a transverse speed of
$200\,\mathrm{km}\,\mathrm{s}^{-1}$. However, the configuration is
highly contrived, and requires the host galaxies of both the carbon
star and lensing object to be unusually faint (see below). The problem
of two faint galaxies can be avoided if the lensing object is located
in either our galaxy or equivalently the host galaxy of the carbon
star, but then the lensing object has to be very massive (several
hundred solar masses). We therefore reject microlensing as playing any
role in explaining SCP\,06F6.

We suggest as a possible scenario that SCP\,06F6 is related to a
supernova-like event of a carbon-rich star.  The peak magnitude of
SCP\,06F6 is comparable to the peak in the luminosity function of core
collapse supernovae. The implied light curve is shown in
Fig.\,\ref{f-lc}, where it is compared to several other supernovae.
As noted by \citet{barbaryetal08-1} the transient is very slow in
comparison to the majority of SN, which reach their peak on time
scales of $\sim 30$ days or less. However, some SN, such as the IIn
SN\,1994Y \citep{hoetal01-1} and SN\,2006gy \citep{smithetal07-2}, do
appear to have broad peaks, albeit with poor sampling of the rise.
Overall the ensemble of SN light curves seem to broadly incorporate
that of SCP\,06F6.  If SCP\,06F6 is related to supernovae, it will
involve a rapid expansion of the envelope. For a peak magnitude
$M_I=-18$, an assumed temperature of 5000\,K, and a redshift of
$z=0.14$, the radius of SCP\,06F6 would be
$\sim3.5\times10^{10}$\,km. Assuming further that the expansion to
this dimension occurred over $\sim100$\,d, an expansion velocity of
$\sim4000\,\mathrm{km\,s^{-1}}$ is implied. Given the already broad
nature of the C$_2$ Swan bands, the velocity gradient across the
visible fraction of the envelope would result only in a relatively
mild broadening, that would not distort the general appearance of the
spectral features, but would just somewhat smooth out the sharp Swan
band-heads seen in carbon stars. This is illustrated in
the bottom panel of Fig.\,\ref{f-spectrum}, where we show the our
template spectrum convolved with a  expansion velocity of
$4000\,\mathrm{km\,s^{-1}}$. 

A puzzling aspect should SCP\,06F6 lie at moderate redshift is the
absence of an apparent host galaxy. The nearest detected object to the
location of SCP\,06F6 is a 6 sigma detection of an object with
$z_{850}=25.8$, 1.5 arc-seconds from the transient position. At
redshift $z=0.14$ this corresponds to an absolute magnitude of $M_Z
\sim -13.2$, which is extremely faint.  Only a small proportion of
stars lie in such low mass galaxies, although examples of star forming
galaxies with comparable absolute magnitude (e.g. IC\,1613;
\citealt{dolphinetal01-1}) can be found. A number of supernovae and
GRBs have occurred in faint ($M_R\ga13$) galaxies
\citep[e.g.][]{levanetal05-1, fruchteretal06-1, drakeetal08-1},
indicating that such an association is not impossible.  The presence
of C$_2$ Swan bands in the spectrum of the SCP\,06F6 requires a carbon
to oxygen ratio of C/O$>1$ by number, such that the carbon is not
locked up in CO, favouring regions of low metallicity where it is
easier for a small amount of carbon production to overwhelm the oxygen
abundance. Given the well known relation between mass (or luminosity)
and metallicity \citep{tremonti04etal-1}, such carbon rich events may
preferentially occur in faint host galaxies. Deep spectroscopy of the
hypothetical nearby galaxy may enable a measure of its redshift, to
test the association with SCP\,06F6 if at $z\simeq0.14$. The X-ray
luminosity of SCP\,06F6 inferred from the \textit{XMM-Newton}
observations,
$L_\mathrm{X}=(5\pm1)\times10^{42}\,\mathrm{ergs\,s^{-1}}$, is much
larger than expected from normal core collapse supernovae
\citep{unoetal02-1, kouveliotouetal04-1}, though the SN\,IIn 1988Z and
2006jd reached
$L_\mathrm{X}\simeq1-2.5\times10^{41},\mathrm{erg\,s^{-1}}$
\citep{fabian+terlevich96-1, immleretal07-1}. It is intriguing that
both the slowly evolving light curve and large X-ray luminosity of
SCP\,06F6 bear similarities to the behaviours observed in SN\,IIn,
despite the radically different spectral appearance. A speculative
scenario is that SCP\,06F6 is associated with the death of a massive
star that underwent mass loss removing the hydrogen envelope, e.g. a
WC Wolf-Rayet star prior to the explosion, creating a carbon-rich
circumstellar environment, or a Iben \& Renzini type\,1.5 supernova
inside a metal-poor AGB star \citep{zijlstra04-1}, which might possess
a carbon-rich envelope at the point of core ignition. Interaction of
the reverse shock with dense circumstellar matter would then also
provide an explanation for the X-ray emission of SCP\,06F6
\citep{chevalier+fransson94-1}. We note in passing that
\citet{thompsonetal08-1} related the luminous transients SN\,2008S and
NGC\,300 to either core-collapse supernovae or bright eruptions of
massive dust-enshrouded stars, possibly carbon stars.

An alternative hypothesis of SCP\,06F6 is that it is not due to
stellar collapse but rather to the tidal disruption of a carbon rich
star by a black hole. Such tidal disruption events may occur in the
core of normal galaxies when stars approach the central black hole
\citep{rees88-1}; or further out in hosts via intermediate mass black
holes \citep[e.g.][]{rosswogetal09-1}. These events can potentially
create the light curve shape (long duration flare) and approximate
luminosity seen in SCP\,06F6, for example object D3-3 in
\citet{gezarietal08-1} reaches a peak magnitude of $M_\mathrm{g}\sim
-19$ and has an transient duration of several hundred days. The X-ray
luminosity of SCP\,06F6 is comparable to that of the candidate tidal
disruption events identified in the \textit{XMM-Newton} Slew Survey
\citep{esquejetal07-1}. This interpretation has the advantage that it
more naturally explains the X-ray luminosity in tandem with that in
the optical. Finally, tidal disruption events may lead to the ejection
of a large fraction of the material of the disrupted star
\citep{ayaletal00-1}, explaining the presence of a large optically
thick envelope.

However, this interpretation is not without problems. 
The lack of any obvious host galaxy to very low luminosities would
imply either a very low black hole mass (if black holes do exist at
the centres of dwarf irregulars) or that the black hole has somehow
been ejected from its host (as has been suggested in a few cases
e.g. \citealt{magainetal05-1, haehneltetal06-1}). These possibilities,
combined with the observation that the disrupted object be a carbon
rich star, rather than a normal main sequence one appear to make the
case for tidal disruption somewhat contrived. Nonetheless, with only
one object, and thus an essentially unconstrained rate and space
density for such events, it remains a possibility.

Any model of this source will need to explain both its unusual
spectral appearance and the implied abundances as well as the high
X-ray luminosity.

\section{Summary}
We have suggested that the unusual transient recently reported by
\citet{barbaryetal08-1}, can be interpreted as being due to a carbon
rich supernova-like event.  We identify the features in the optical
spectrum of SCP\,06F6 as being due to carbon Swan bands, at a redshift
of $z\simeq0.14$. At this redshift the energetics of SCP\,06F6 resemble
those of core collapse supernovae, albeit with a longer than typical
rise time, and higher than typical X-ray luminosity. If correct, this
suggests that the rare collapse of carbon-rich stars can yield
supernovae very different from the bulk populations which are
frequently observed in current transient searches, and further
motivates the next
generation of transient experiments.

\acknowledgements{We thank the Supernova Cosmology Project for making
their reduced spectra of SCP\,06F6 available. This research was 
supported by STFC. The SDSS is managed by the Astrophysical Research
Consortium for the Participating Institutions. We thank the referee
for a number of useful suggestions.}


\end{document}